\documentclass[aps,prl,twocolumn,showpacs,amsmath,amsmath,amssymb,amsfonts,superscriptaddress]{revtex4-1}
\usepackage{graphicx}
\usepackage{dcolumn}
\usepackage{bm}
\usepackage{color}
\usepackage{multirow}

\begin{document}

\title{Supersolidity of lattice Bosons immersed in strongly correlated Rydberg dressed atoms}
\author{Yongqiang Li}
\affiliation{Department of Physics, National University of Defense Technology, Changsha 410073, P. R. China}
\author{Andreas Gei{\ss}ler}
\affiliation{Institut f\"ur Theoretische Physik, Goethe-Universit\"at, 60438 Frankfurt/Main, Germany}
\author{Walter Hofstetter}
\affiliation{Institut f\"ur Theoretische Physik, Goethe-Universit\"at, 60438 Frankfurt/Main, Germany}
\author{Weibin Li}
\affiliation{School of Physics and Astronomy, University of Nottingham, Nottingham NG7 2RD, UK}
\affiliation{Centre for the Mathematics and Theoretical Physics of Quantum Non-equilibrium Systems, University of Nottingham, Nottingham NG7 2RD, UK}

\date{\today}
\begin{abstract}
Recent experiments have illustrated that long range two-body interactions can be induced by laser coupling atoms to highly excited Rydberg states. Stimulated by this achievement, we study supersolidity of lattice bosons in an experimentally relevant situation. In our setup, we consider two-component atoms on a square lattice, where one species is weakly dressed to an electronically high-lying (Rydberg) state, generating a tunable, soft-core shape long-range interaction. Interactions between atoms of the second species and between the two species are characterized by local inter- and intra-species interactions. Using a dynamical mean-field calculation, we find that interspecies onsite interactions can stabilize a pronounced region of supersolid phases. This is characterized by two distinctive types of supersolids, where the bare species forms supersolid phases that are immersed in strongly correlated quantum phases, i.e. a crystalline solid or supersolid of the dressed atoms. We show that the interspecies interaction leads to a roton-like instability in the bare species and therefore is crucially important to the supersolid formation. We provide a detailed calculation of the interaction potential to show how our results can be explored under current experimental conditions.
\end{abstract}


\maketitle

A supersolid is a translational symmetry breaking superfluid occurring in a solid. It was predicted to exist in bulk helium over forty years ago~\cite{supersolid}, but its observation has remained a challenge~\cite{Kim}. To reach supersolidity, one typically relies on long-range two-body interactions to break the translational invariance of a homogeneous system. Recent experiments have observed supersolid orders where translational symmetry is broken by cavity photon assisted~\cite{cavity} or spin-orbit coupling enabled~\cite{SOC} momentum transfer. \begin{figure}[ht!]
	\includegraphics*[width=0.905 \columnwidth]{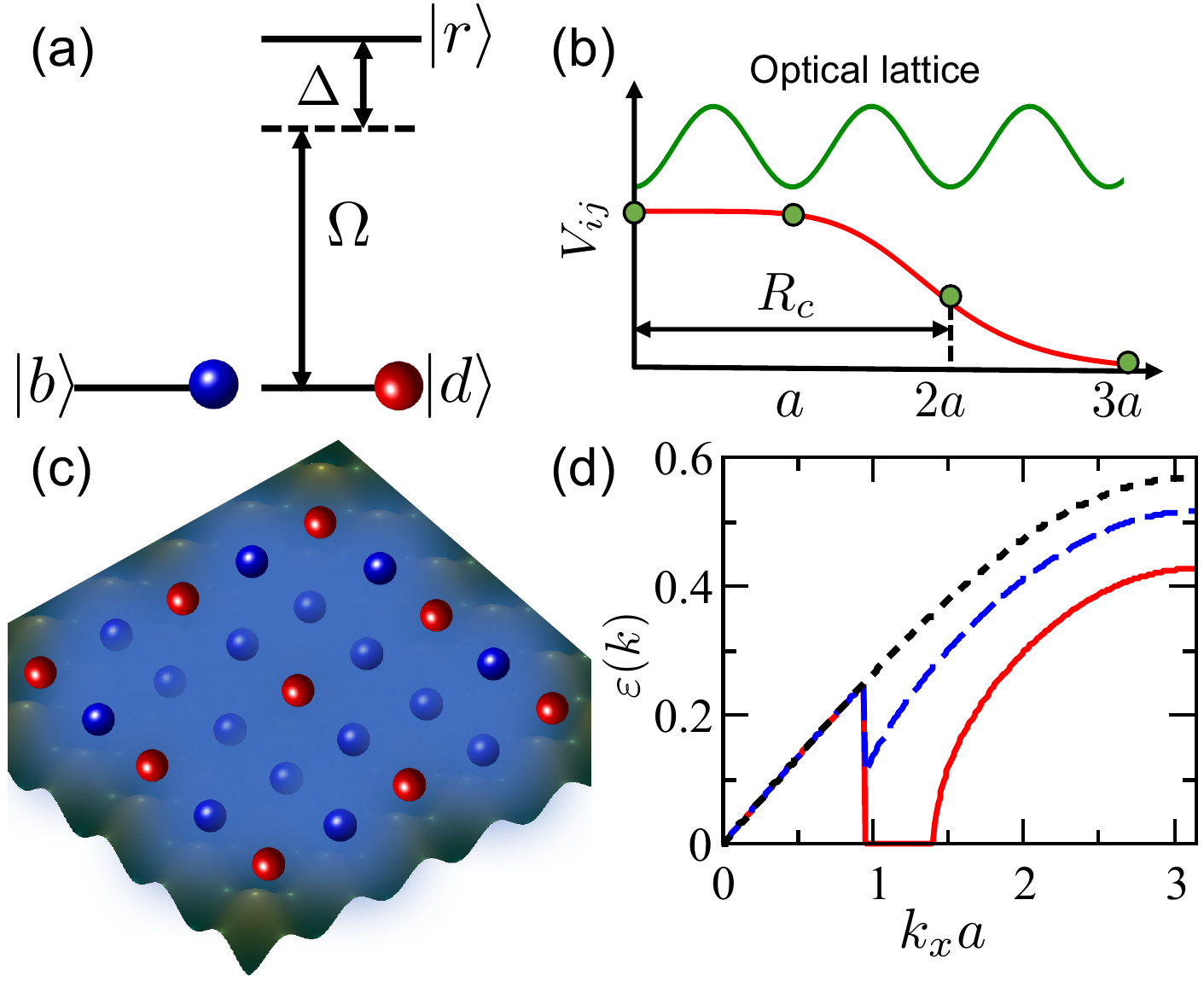}
	\vspace{-3.5mm}
	\caption{(Color online) (a) Two electronic ground states $|b\rangle$ (blue) and $|d\rangle$ (red) and a Rydberg state $|r\rangle$ are considered. An off-resonant laser (with Rabi frequency $\Omega$ and detuning $\Delta$) weakly couples the state $|d\rangle$ to $|r\rangle$. (b) The soft-core shape interaction potential $V_{ij}$ (red) between atoms in the Rydberg dressed state $|d\rangle$. The soft-core radius $R_c$ can be larger than the lattice spacing $a$. Here $R_c=2a$ is shown. (c) SS of the bare state when dressed atoms are in an ordered density wave (DW). (d) Roton instability of the bare species. The Bogoliubov dispersion relation (along the $k_x$ axis)  of phonons is significantly modified by the interspecies interaction. A roton-like instability emerges when the interspecies interaction $U_{bd}$ is increased, indicating that the ground state phase changes from a homogeneous superfluid to supersolid. In the figure we show  $U_{bd}/U=0$ (dotted), $U_{bd}/U=0.45$ (dashed) and $U_{bd}/U=1$ (solid). Other parameters are $k_y=0$, $V/U=0.4$ and $t/U=0.04$. See text for details. }
	\label{fig1}
\end{figure}
To achieve supersolids induced purely by two-body interactions, enormous efforts have been spent on polar molecules~\cite{Ye, Ye_theory}, magnetic~\cite{Cy_Dr_Ey} and Rydberg atoms~\cite{single_component,single_eBH}, due to the available long-range atom-atom interaction as well as high precision control over their internal and motional states. However, a current challenge is that theoretical proposals typically examine regimes that are difficult to achieve experimentally.



\begin{figure*}[t]
	\vspace{-10mm}
	\includegraphics*[width=.9 \linewidth]{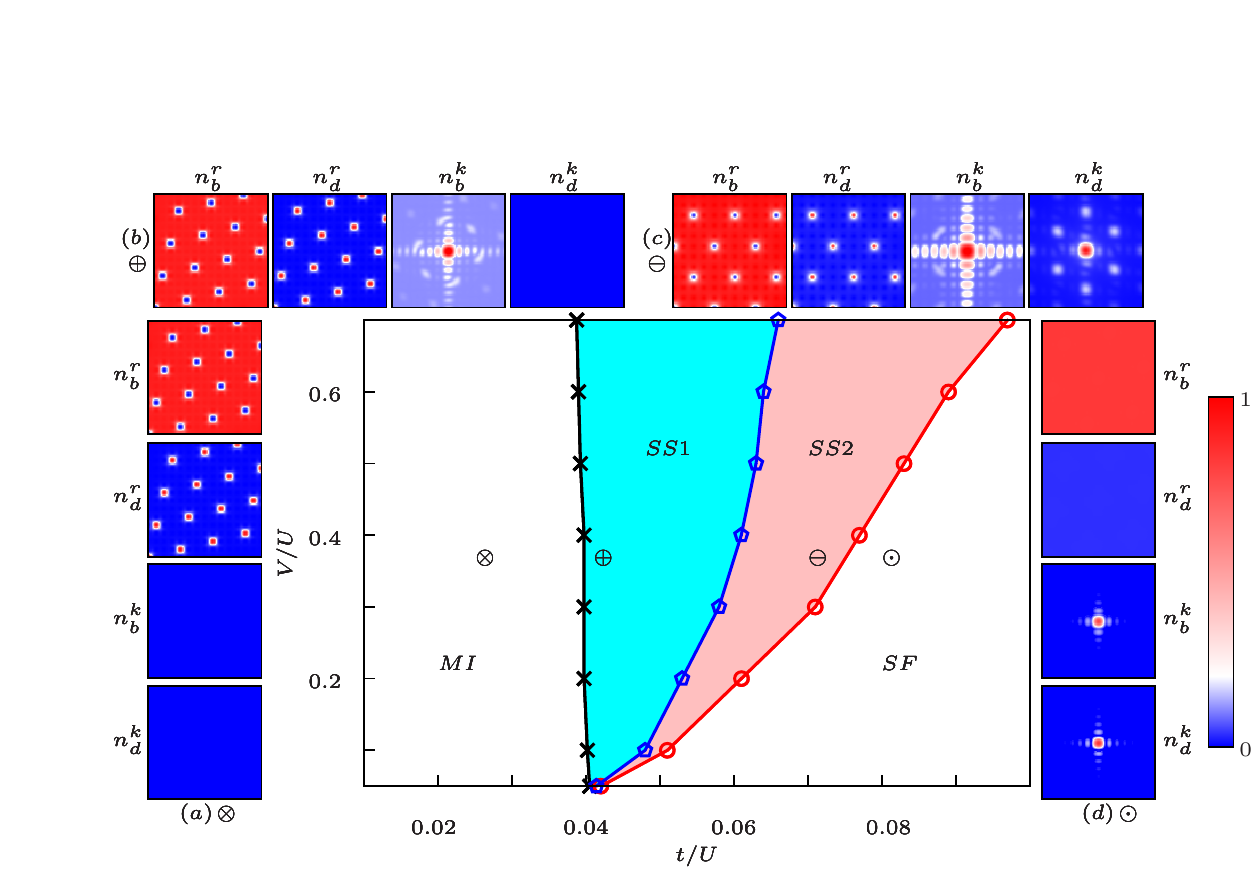}
	\vspace{-5.5mm}
	\caption{(Color online) Phase diagram of a mixture of ground-state component $b$ and Rydberg dressed species $d$ on a square lattice in terms of hopping amplitude $t$ and Rydberg dressed interaction strength $V$. There are four stable phases in the diagram: Mott insulator (MI) with spatially uniform total local density and crystalline density order for each species, homogeneous superfluid (SF), and two types of supersolids (SS1 with Rydberg dressed species being in the crystalline phase, and SS2 with both species being in the supersolid). Other parameters are $U_{bd}=U$ and $n^r_b+n^r_d=1$. (a)-(d): Real-space density $n^r_{b,d}$ and quasi-momentum-space density $n^k_{b,d}$ distributions of different phases, with lattice sizes being the square of the area of the unit cell of the Rydberg dressed species [${\rm MI}, N_{\rm lat} = 15\times15 \,(\otimes)$; ${\rm SS1}, N_{\rm lat} = 15\times15 \,(\oplus)$; ${\rm SS2}, N_{\rm lat} = 12\times12 \,(\ominus)$; and ${\rm SF}, N_{\rm lat} = 24\times24\,(\odot)$], as shown by the markers in the main figure. }
	\label{phase_diagram}
\end{figure*}
In this work, we study supersolids of a two-species bosonic mixture on a two-dimensional (2D) square lattice, where one of the species is weakly coupled to an electronically high-lying (Rydberg) state by an off-resonant laser (the level scheme is depicted in Fig.~\ref{fig1}a). Uniquely, this setting is recently realized experimentally at Munich~\cite{Rydberg} in the study of Rydberg dressed spin dynamics~\cite{spin_lattice}. The coupling laser induces strong and long-range interactions between Rydberg dressed atoms on distances well beyond typical lattice-site spacings (see Fig.~\ref{fig1}b), whose strength and sign can be controlled by the laser (i.e. detuning and Rabi frequencies) and the choices of Rydberg states~\cite{dressed}. The resulting Bose-Hubbard model features a long-range interaction between dressed atoms while interactions between atoms of the two different species and of the bare species are short ranged.

Employing \emph{real-space} bosonic dynamical mean-field theory (RBDMFT), we find that the system undergoes a series of many-body phases, including Mott insulator (MI), ordered density wave (DW), supersolid (SS) and superfluid (SF) phases. A key result is that the interspecies interaction enables supersolid phases of the bare species in regions where the dressed atoms are in DW or SS phases (an example for a DW is depicted in Fig.~\ref{fig1}c).
Using Bogoliubov theory, we reveal that a roton-like instability emerges due to the interspecies interaction (see Fig.~\ref{fig1}d), which signifies a SF to SS transition~\cite{SF_SS}. Our results open a new route to enhance the formation of SS phases through the Rydberg dressing in two-component atomic gases.

{\it The Hamiltonian--} In sufficiently deep lattices, our setting is described by a single band, two-component Bose-Hubbard model,
\begin{eqnarray}\label{Hamil}
\hat{H} = &-& \sum_{\langle ij \rangle,\sigma}t_\sigma(\hat{b}^\dagger_{i\sigma}\hat{b}_{j\sigma} + {\rm H.c.})
+ \sum_{i<j}V_{ij}\hat{n}_{id}\hat{n}_{jd} - \sum_{i} \hat{H}_{i}\nonumber,
\end{eqnarray}
where the single site Hamiltonian $\hat{H}_{i} = \frac12 \sum_{\sigma\sigma^\prime} U_{\sigma\sigma^\prime} \hat{n}_{i\sigma}(\hat{n}_{i\sigma^\prime}-\delta_{\sigma\sigma^\prime}) - \sum_{\sigma} \mu_\sigma \hat{n}_{i\sigma}$.
$\langle i,j\rangle$ represents the nearest neighbour sites $i,j$. Index $\sigma (\sigma^\prime)=b, d$ denotes bare, and dressed states, respectively. $\hat{b}^\dagger_{i\sigma}$ ($\hat{b}_{i\sigma}$) and $\hat{n}_{i\nu}=\hat{b}^\dagger_{i\nu}\hat{b}_{i\nu}$ are the bosonic creation (annihilation) operator for species $\sigma$ and atomic density at site $i$. $t$ and $\mu_\sigma$ determine the hopping rate and chemical potential for the two bosonic species. We assume the hopping rates are identical for both species~\cite{identical_hopping}. $U_{\sigma\sigma^\prime}$ denotes the inter- and intra-species short-range (onsite) interactions, which can be tuned via e.g. Feshbach resonances~\cite{Feshbach} or state-dependent optical lattices~\cite{state_lattice}. The long-range interaction between site $i$ and $j$ is $V_{ij}\equiv V / [(a/R_c)^6(i-j)^6 + 1]$, where $V=\tilde{C}_6/R_c^6$ characterises the long-range interaction at a distance $R_c$. $\tilde{C}_6$, $R_c$ and $a$ are the effective dispersion coefficient, soft-core radius, and lattice constant, respectively. In the following, we choose the intraspecies short-range interaction $U_{b,d}\equiv U$, which also sets the unit of energy. Details of these parameters will be given towards the end of the paper.

To determine the ground state phases, we use RBDMFT to capture both higher order quantum fluctuations, strong correlations and arbitrary long-range order in a unified framework~\cite{RBDMFT,BDMFT_2009}. It provides a nonperturbative description of many-body systems in two and three spatial dimensions (the method is discussed in the supplementary material.). In the calculations, we typically consider the lattice size as large as $N_{\rm lat}=48\times48$ sites and an experimentally relevant  soft-core radius $R_c=3a$~\cite{dressed}. The superfluidity is characterised by  the condensate order parameter $\phi_\sigma \equiv \langle \hat{b}_\sigma \rangle$, and crystalline order by the real-space density distribution $n_{i\sigma}=\langle \hat{n}_{i\sigma}\rangle$ and total density $n_i\equiv n_{ib}+n_{id}$.  The coexistence of both condensate and crystalline order parameters gives the supersolid phase. Note that a similar model using dipolar gases has been numerically investigated using a mean-field Gutzwiller approach and by considering only the nearest-neighbor part of the dipolar interactions~\cite{NN}. In our calculations, we take into account the whole range of the interaction potential (see appendix for a comparison of the two systems).

{\it Many-body ground state phase diagram---} The main results are summarized in the phase diagram shown in Fig.~\ref{phase_diagram}. Depending on the parameters, the two-component system can have five different phases, i.e. Mott insulator (MI), ordered density wave (DW), two types of supersolid (SS1 and SS2), and superfluid (SF). In the following, we will discuss features of these phases for unit filling $n_{jd} +n_{jb}=1$ (see appendix for results at other fillings).

\begin{figure}[h]
	\includegraphics*[width=0.9 \columnwidth]{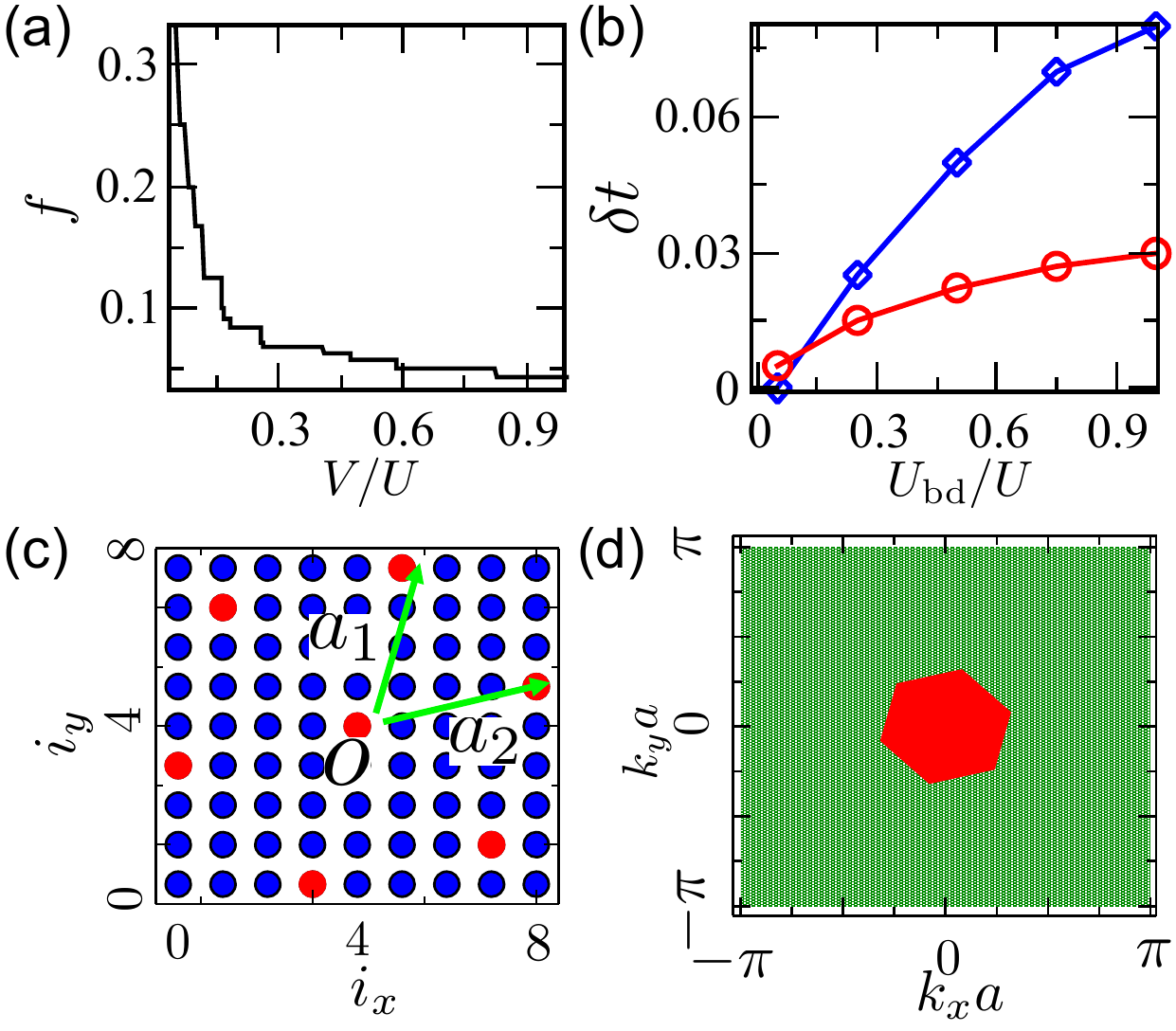}
\vspace{-3.5mm}
	\caption{(Color online) (a) Devil's staircase pattern of the filling fraction $f=\sum_i n_{id}/N_{\rm lat}$ for the Rydberg dressed species in the zero-hopping limit.  (b) Width of supersolid phase SS1 (blue) and SS2 (red) $\delta t \equiv t_{\rm c1}-t_{\rm c2}$ as a function of interspecies interaction $U_{bd}/U$ for Rydberg dressed interaction $V/U=0.1$, where $t_{\rm c1,2}$ denotes the critical value of the hopping amplitude of the upper/lower phase boundary of each phase shown in Fig.~\ref{phase_diagram}. (c) Density distribution of the dressed (red) and bare species (blue). The dressed atoms form an oblique lattice with lattice vector $a_1$ and $a_2$. This structure corresponds to the configuration illustrated in Fig.~\ref{phase_diagram}b. (d) The first Brillouin zone of the optical lattice (green) and oblique lattice (red) of the dressed atom. As the lattice vector $|a_j|>a$ ($j=1,2$), the size and shape of the first Brillouin zone of the dressed atoms differ significantly from the square reciprocal lattice of the optical lattice potential. }
	\label{bogoliubov}
\end{figure}

We start with the so-called strong coupling limit when $U_{\sigma\sigma^\prime} \gg t$, where the 2D system favours MI phases with uniform total particle densities. Crystalline orders in the MI region can be changed by varying the two-body interactions (i.e. $V/U$). One example is depicted in Fig.~\ref{phase_diagram}a, which shows relative densities and crystalline structures. Furthermore, when one increases $V/U$ continuously, the filling fractions $f_d\equiv \sum_i n_{id}/N_{\rm lat}$ of the dressed species can form a devil's staircase structure (Fig.~\ref{bogoliubov}a). An open question here is whether the staircase in this 2D system is complete. In 1D lattice systems, the devil's staircase and its completeness~\cite{one_dimension} have been extensively studied~\cite{staircase}. Moreover, there are very small regions occupied by DW phases (with a non-uniform total density). Due to that, the corresponding discussion will be given in the supplementary material.



When the hopping rate increases, we observe a pronounced region of supersolids. The bare state first enters the supersolid phase (SS1) from an insulating phase, while the dressed species is still crystallized in this case (one example is depicted in Fig.~\ref{phase_diagram}b). Further increasing $t$, both species are in supersolid phases (SS2), as shown in Fig.~\ref{phase_diagram}c, where non-zero peaks appear for both species in addition to zero-momentum condensate, indicating the coexistence of non-trivial diagonal long-range order and off-diagonal long-range order associated with phase coherence. A large supersolid region indicates a higher chance for directly observing these phases in realistic experiments, compared to the single-species case~\cite{single_component}.

One typically would not expect such supersolids as the bare species alone can only form superfluid and MI phases due to the short range two-body interactions~\cite{two_component_SF_MI}. The underlying mechanism is that the flow of the bare species is suppressed by the crystalline distribution of the dressed species via the interspecies interaction. As a result, the widths of the SS1 and SS2 phases will strongly depend on the interspecies interaction $U_{\text{bd}}$. The numerical result in Fig.~\ref{bogoliubov}b shows that indeed the two SS phases shrink as $U_{bd}$ decreases. The SS1 phase eventually disappears for sufficiently small $U_\text{bd}$.

For even larger hopping rate $t$, both species are in SF phases, which are characterized by nonzero SF order parameters. Different from the SS, spatial densities of both species become homogeneous in the SF phases.

{\it Supersolidity mechanism of the bare species---} In the rest of the work, we will develop a Bogoliubov mean-field theory to understand how the interspecies interaction enables the bare species to form SS phases. Our discussion will focus on the SS1 phase, where the dressed species is a DW. This allows us to write down wave functions $|{\rm{DW}}_d\rangle$ of the DW according to the crystalline structure. We also assume that the total wave function in the ground state can be decoupled as $|\Psi_{\rm{g}}\rangle\approx |{\rm{DW}}_d\rangle \otimes |\Psi_b\rangle$, where $|\Psi_b\rangle$ is the wave function of the bare component. Then we can derive an effective Hamiltonian for the bare species by tracing out the dressed atom part, i.e. $\hat{H}_{\rm{e}} = \langle {\rm{DW}}_d|\hat{H}|{\rm{DW}}_d\rangle$. Explicitly the effective Hamiltonian reads,
\begin{eqnarray}
\hat{H}_{\rm{e}}= &-& \sum_{\langle ij \rangle}t(\hat{b}^\dagger_{i}\hat{b}_{j} + {\rm H.c.})
+\frac{U}{2} \sum_{i} \hat{n}_{i}(\hat{n}_{i}-1) \nonumber \\
&-&  \sum_{i} \mu \hat{n}_{i}+U_{bd}\sum_{\{j\}} \hat{n}_{j},\nonumber
\end{eqnarray}
where ${\{j\}}$ denotes lattice sites occupied by dressed atoms. For convenience, we have omitted the index $b$ of the bare species. The last term gives the interspecies interaction, where the mean particle number per site of the dressed atoms $n_d=1$ has been used explicitly. A constant term,  $C=\langle {\rm{DW}}_d|\sum_{i<j}V_{ij}\hat{n}_{id}\hat{n}_{jd}|{\rm{DW}}_d\rangle$ characterizing the long-range interaction energy, is neglected in the effective Hamiltonian.

The interaction with the dressed atoms (the last term in the effective Hamiltonian) introduces a new spatially periodic structure to the bare species, in addition to the optical lattice. As an example, we consider parameters corresponding to Fig.~\ref{phase_diagram}b. Here, the dressed atoms form an oblique lattice, see Fig.~\ref{bogoliubov}c for a cartoon picture of the 2D structure. The  primitive cell of the new oblique lattice is apparently larger than the original lattice. In this example, the primitive lattice vectors are $a_1=(1,\,4)$ and $a_2=(4,\,1)$, with which we obtain the area of the primitive lattice $A=|a_1\times a_2|=15$, while the area of the optical lattice is 1. In turn, the corresponding reciprocal lattice is smaller than that of the optical lattice. To illustrate this, we plot the first Brillouin zone of the two lattices in Fig.~\ref{bogoliubov}b. Apparently they overlap only in a small central area (low momentum regions).

As a result, phonon excitations for momentum components in and out of the overlap region will be very different. To show this, we calculate the Bogoliubov dispersion relation of the effective Hamiltonian. In the low momentum region (where the two Brillouin zones overlap), $E_k^{\text{i}}=\sqrt{\varepsilon_k^2 + 2\bar{n}_bU\varepsilon_k}$ with $\varepsilon_k=-2t(\cos k_x a+ \cos k_y a-2)$. Outside this region, the dispersion becomes $E_k^{\text{o}}=\sqrt{\left(\varepsilon_k- \bar{n}_dU_{bd} \right)^2 + 2\bar{n}_bU\left(\varepsilon_k-  \bar{n}_dU_{bd}\right)}$. Here $\bar{n}_b$ ($\bar{n}_d$) are the mean population of the bare (dressed) component.
Consequently, the dispersion is not continuous any more at the boundary of the Brillouin zone of the oblique lattice. The dispersion relation becomes complex when $U_{bd}>2t\bar{n}_d (2-\cos k^{(b)}_xa - \cos k^{(b)}_ya)$ where $k^{(b)}_x$ and $k^{(b)}_y$ are momenta at the boundary.  In Fig.~\ref{fig1}d, we plot the dispersion relation along the $k_x$ axis by varying the interspecies interaction $U_{bd}$, where the mode frequency becomes complex at $U_{bd}=U$. This so-called roton-like instability~\cite{SF_SS} here indicates that the emergence of supersolids is indeed induced by the strong interspecies interaction. Note that the mechanism here is different from SS phases induced by geometrically dependent hopping found in frustrated lattices~\cite{frustration}.

{\it Interaction potentials of Rydberg dressed atoms---} The level structure used in the Rydberg dressing is shown in Fig.~\ref{fig1}a. The species $|d\rangle$ is coupled to a Rydberg state by an off-resonant laser with Rabi frequency $\Omega$ and detuning $\Delta$. Interactions between Rydberg atoms are of van der Waals type $V_{\text{r}}=C_6/r^6$, where $C_6$ is the respective dispersion coefficient. The Rydberg dressing gives the soft-core interaction $V_{ij}$ where the effective dispersion coefficient $\tilde{C}_6=(\Omega/\Delta)^4C_6$ and soft-core radius $R_c=(C_6/2\Delta)^{1/6}$. $R_c$ varies with the Rydberg states and detuning. For example, one can choose the Rydberg 36S state of $^{87}$Rb atoms ($C_6=241.6\, \text{MHz}\times \mu m^6$) and lattice constant $a=532$ nm. When $\Delta = 7$ MHz,  we obtain $R_c\approx 3a$. With this fixed detuning $\Delta$, the strength of the soft-core interaction is now controlled by the Rabi frequency $\Omega$.

\begin{figure}[h]
	\includegraphics*[width=0.9 \columnwidth]{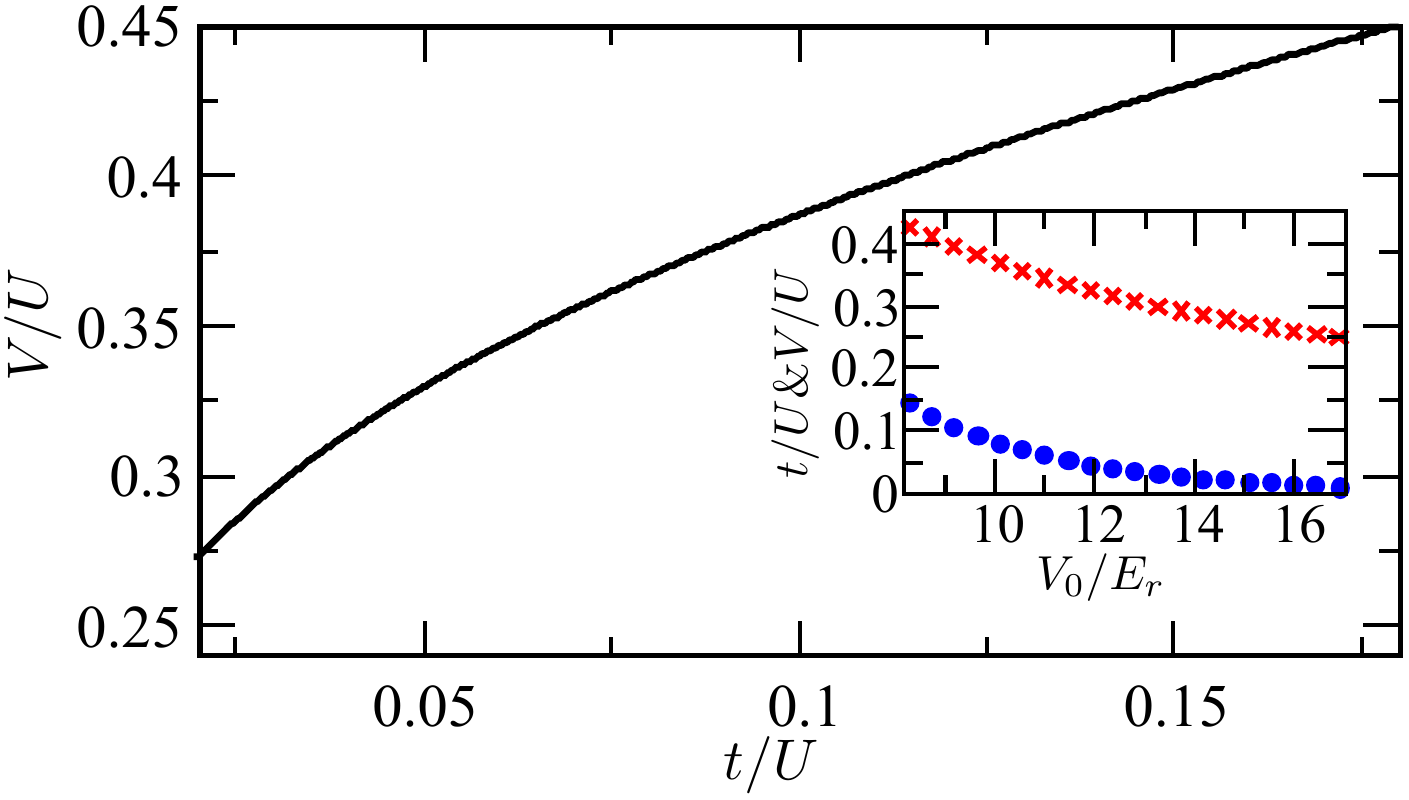}
\vspace{-3.5mm}
	\caption{(Color online) The hopping rate $t$ and onsite interaction $U$ depend on the lattice depth $V_0/E_r$. Increasing $V_0/E_r$, one can observe the phases discussed in the main text. The inset shows changes of $t/U$ ($\color{blue}\bullet$) and $V/U$ ($\color{red}\times$) individually as a function of $V_0/E_r$. Here we consider the Rydberg state 36S of $^{87}$Rb atoms. Other parameters are $\lambda$=1064 nm, $a_s=5.2$ nm, $C_6$=241.6 MHz$/\mu$m$^6$, $\Delta=7$ MHz and $\Omega=0.44$ MHz.}
	\label{parameter}
\end{figure}

To probe different phases shown in Fig.~\ref{phase_diagram}, one needs to change the parameters $V$, $U$ and $t$ together or separately over certain ranges. One simple way to achieve this is to tune the lattice potential depth $V_{0}/E_r$. In  optical lattices, the onsite interaction $U$ depends on the lattice depth through $U= \sqrt{8/\pi}ka_sE_r(V_0/E_r)^{3/4}$ and the hopping rate $t$ through $t=4/\sqrt{\pi} E_r(V_0/E_r)^{3/4}\exp[-2(V_0/E_r)^{1/2}]$~\cite{interaction_hopping}, where $k=2\pi/\lambda$,  $E_r=h^2/2m\lambda^2$, $\lambda=2a$ and $a_s$ are the wave number, recoil energy, wavelength of the lattice potential and s-wave scattering length, respectively. Upon varying $V_0/E_r$ and fixing the other parameters, the ratios $t/U$ and $V/U$ change continuously. One example is shown in Fig.~\ref{parameter}. One can see that the parameters cross the main phases discussed in this paper.

In conclusion, we have investigated crystalline phases of ultracold binary bosonic gases on a square lattice, with one species possessing a non-local interaction induced by Rydberg dressing.  We found two types of supersolid phases that are robust and occupy large parameter regions at zero temperature. We showed that the supersolid phases of the bare species are stabilized by the interspecies interaction. The existence of the different phases predicted here could be directly observed by quantum gas microscopy with single-site resolution~\cite{single_site, M. Greiner_2009, I. Bloch_2010} or through measuring noise correlations~\cite{noise}.
Our results demonstrate rich features of the Bose-Bose mixture with long-range interactions, and indicate that this system is well suited for exploring supersolidity in upcoming experiments. As the crystalline structure (see Fig.~\ref{bogoliubov}a) can be changed in the insulating region by tuning $V/U$, we expect that supersolid phases with tunable density patterns can be explored as well.

We acknowledge J.-M. Yuan, Z.-X. Zhao and Rejish Nath for useful discussions. This work was supported by the National Natural Science
Foundation of China under Grants No. 11304386 and No. 11104350. WL acknowledges support from the UKIERI-UGC Thematic Partnership No. IND/CONT/G/16-17/73 and EPSRC Grant No. EP/M014266/1. AG and WH acknowledge support from the Deutsche Forschungsgemeinschaft via DFG SPP 1929 GiRyd
and the high-performance computing center LOEWE-CSC.

\clearpage

\begin{widetext}
\section{Supplementary Material}
\renewcommand{\theequation}{S\arabic{equation}}
\renewcommand{\thefigure}{S\arabic{figure}}
\renewcommand{\bibnumfmt}[1]{[S#1]}
\renewcommand{\citenumfont}[1]{S#1}
\setcounter{equation}{0}
\setcounter{figure}{0}

\section{Method}
\subsection{RBDMFT equations}
In deriving the effective action, we consider the limit of a high but finite dimensional optical lattice, and use the cavity method~\cite{georges96, Byczuk_2008} to derive self-consistency equations within RBDMFT. In a more formal language, first we map the Hamiltonian onto a set of individual single-site problems each of which is described by a local effective action~\cite{LYQ}:
\begin{small}
\begin{eqnarray}\label{eff_action}
	S^{(i)}_\text{imp} &=& -\int_0^\beta \hspace{-0.2cm} d \tau d\tau' \sum_{\sigma\sigma'} \Bigg( \hspace{-0.1cm} \begin{array}{c} b^{(i)}_{0,\sigma}(\tau)^* \quad b^{(i)}_{0,\sigma} (\tau) \end{array}\hspace{-0.1cm} \Bigg)^{\hspace{-0.1cm}}\boldsymbol{\mathcal{G}}^{(i)}_{0,\sigma\sigma'}(\tau-\tau')^{-1} 	
	\Bigg(\begin{array}{c} \vspace{0.2cm} b^{(i)}_{0,\sigma'} (\tau') \\
	b^{(i)}_{0,\sigma'} (\tau')^* \end{array} \hspace{-0.1cm}\Bigg) +	\\
	&& \hspace{-0.2cm} \quad\int_0^\beta  d\tau   \left\{ \frac{1}{2}U_{\sigma\sigma^\prime}\; n^{(i)}_{0,\sigma}(\tau) \,\Big( n^{(i)}_{0,\sigma^\prime}(\tau) - \delta_{\sigma\sigma^\prime} \Big) + \sum_{j(j \neq 0)} V_{0j}n^{(i)}_{0,d}(\tau) n^{(i)}_{j,d}(\tau)
	- \sum_{\langle 0j\rangle,\sigma } t_{\sigma} \Big( b^{(i)}_{0,\sigma}(\tau)^* \phi^{(i)}_{j,\sigma}(\tau)
	+ b^{(i)}_{0,\sigma}(\tau) \phi^{(i)}_{j,\sigma}(\tau)^* \Big) \right\}. \nonumber
	\end{eqnarray}
\end{small}
Here we have defined the local Weiss Green's function,
\begin{eqnarray}
\hspace{-5mm}\boldsymbol{\mathcal{G}}^{-1}_{0,\sigma\sigma'}(\tau-\tau') \equiv
- \left(\begin{array}{cc} \hspace{-0.1cm}
				(\partial_{\tau'}-\mu_\sigma)\delta_{\sigma \sigma'}+ t^2
\hspace{-0.25cm} \sum \limits_{\langle 0i\rangle,\langle 0j\rangle} \hspace{-0.25cm}
G_{\sigma \sigma', ij}^1 (\tau, \tau')
			&  t^2  \hspace{-0.25cm} \sum \limits_{\langle
0i\rangle,\langle 0j\rangle} \hspace{-0.25cm}   G^2_{\sigma\sigma', ij}(\tau, \tau') \\
			 t^2  \hspace{-0.25cm} \sum \limits_{\langle
0i\rangle,\langle 0j\rangle} \hspace{-0.25cm} {G^2_{\sigma\sigma', ij}}^*(\tau', \tau)
	 & (-\partial_{\tau'}-\mu_\sigma)\delta_{\sigma \sigma'}+ t^2   \hspace{-0.25cm}
\sum \limits_{\langle 0i\rangle,\langle 0j\rangle} \hspace{-0.25cm} G_{\sigma \sigma', ij}^1
(\tau', \tau)
			 \hspace{-0.1cm} \end{array}\right) \hspace{-0.15cm},	
\end{eqnarray}
and introduced
\begin{equation}
\phi^{}_{i,\sigma}(\tau) \equiv \langle b_{i, \sigma} (\tau)
\rangle_0
\end{equation}
as the superfluid order parameters, and
\begin{eqnarray}
\hspace{-0.5cm}G_{\sigma\sigma', ij}^1 (\tau, \tau')\hspace{-0.2cm}&\ \equiv\ &\hspace{-0.2cm}- \langle b_{i, \sigma} (\tau) b_{j, \sigma'}^* (\tau')
\rangle_0 + \phi_{i, \sigma'} (\tau) \phi_{j, \sigma}^* (\tau'), \\
\hspace{-0.5cm}G_{\sigma \sigma', ij}^2 (\tau, \tau')\hspace{-0.2cm}&\ \equiv\ &\hspace{-0.2cm}- \langle b_{i, \sigma} (\tau) b_{j, \sigma'} (\tau')
\rangle_0 + \phi_{i, \sigma'} (\tau) \phi_{j, \sigma} (\tau')
\end{eqnarray}
as the diagonal and off-diagonal parts of the connected Green's functions, respectively, where $\langle \ldots \rangle_0$ denotes the expectation value in the cavity system (without the impurity site)~\cite{LYQ, Walter}.

\subsection{Anderson impurity model}
The most difficult step in the procedure discussed above is to find a solver for the effective action. However, one cannot do this analytically. To obtain RBDMFT equations, it is better to return back to the Hamiltonian representation. Here, each
of the local effective actions~(\ref{eff_action}) is represented by an Anderson impurity Hamiltonian
\begin{eqnarray}
\hat{H}_A =& - & \sum_{\langle 0 j\rangle \sigma} t_\sigma \big(\phi^*_{j,\sigma} \hat{b}_{0,\sigma} + {\rm h.c.} \big) + \frac{1}{2}\sum_{\sigma\sigma^\prime}U_{\sigma\sigma^\prime} \hat{n}_{0,\sigma}( \hat{n}_{0,\sigma^\prime} - \delta_{\sigma\sigma^\prime}) + \sum_{j(j\neq 0)} V_{j0}\langle {\hat n}_{j,d} \rangle \hat{n}_{0,d} - \sum_\sigma \mu_{0,\sigma} \hat{n}_{0,\sigma} \nonumber \\
         &+&  \sum_{l}  \epsilon_l \hat{a}^\dagger_l\hat{a}_l + \sum_{l,\sigma} \Big( V_{\sigma,l} \hat{a}^\dagger_l\hat{b}_{0,\sigma} + W_{\sigma,l} \hat{a}_l\hat{b}_{0,\sigma} + {\rm h.c.} \Big),
\end{eqnarray}\label{Anderson}
where the chemical potential and interaction term are directly inherited from the Hubbard
Hamiltonian. The bath of condensed bosons is represented by the Gutzwiller term with
superfluid order parameters $\phi_\sigma$ for each component. The bath of normal bosons is
described by a finite number of orbitals with creation operators $\hat{a}^\dagger_l$ and energies
$\epsilon_l$, where these orbitals are coupled to the impurity via normal-hopping amplitudes
$V_{\sigma, l}$ and anomalous-hopping amplitudes $W_{\sigma, l}$. The anomalous hopping terms are
needed to generate the off-diagonal elements of the hybridization function. Note here that in the high-dimensional limit inter-site interactions only contribute to the Hartree level~\cite{Hartree}. In other words, the Hartree term of the inter-site interaction will dominate as the spatial dimension of the system increases. This motivates us to keep only the Hartree contribution of the inter-site interaction in our simulations as an approximation to the original Hamiltonian, i.e.
\begin{eqnarray}
\frac12 \sum_{i\neq j}V_{ij}\hat{n}_{i,d}\hat{n}_{j,d} \approx \sum_{i\neq j} V_{ij} \langle \hat{n}_{i,d} \rangle (\hat{n}_{j,d} - \frac12 \langle \hat{n}_{i,d} \rangle )
\end{eqnarray}

We now turn to the solution of the impurity model. In practice, we start with an initial set of Anderson paramters and local bosonic superfluid order parameters $\phi_{j,\nu}(\tau)$. The Anderson Hamiltonian can straightforwardly be implemented in the Fock basis, and the corresponding solution can be achieved by exact diagonalization (ED) of DMFT~\cite{M. Caffarel_1994, georges96}. After diagonalization, the local Green's function, which includes all the information about the bath, can be obtained from the eigenstates and eigenenergies
in the Lehmann-representation
\begin{eqnarray}
G_{\rm imp,\sigma \sigma'}^1 (i \omega_n) &=& \frac{1}{Z} \sum_{mn} \langle m | \hat b_\sigma | n\rangle \langle n | \hat b_{\sigma'}^\dagger | m \rangle \frac{e^{- \beta E_n} - e^{-\beta E_m}}{E_n - E_m + i \hbar \omega_n} + \beta \phi_\sigma \phi^\ast_{\sigma'} \\
G_{\rm imp,\sigma \sigma'}^2 (i \omega_n) &=& \frac{1}{Z} \sum_{mn} \langle m | \hat b_\sigma | n\rangle \langle n | \hat b_{\sigma'} | m \rangle \frac{e^{- \beta E_n} - e^{-\beta E_m}}{E_n - E_m + i \hbar \omega_n} + \beta \phi_\sigma \phi_{\sigma'}.
\end{eqnarray}

Integrating out the orbitals leads to the same effective action as in Eq.~(\ref{eff_action}), if the following
identification is made
\begin{eqnarray}
\boldsymbol{\Delta}_{\sigma\sigma'} (i\omega_n)  & \equiv & t^2 {\sum_{\langle0i\rangle,\langle 0j\rangle}}\mathbf G_{\sigma\sigma',ij}(i\omega_n),
\end{eqnarray}
where $\mathbf G_{\sigma\sigma',ij}(i\omega_n)$ is the inverse Fourier transformation of the Weiss Green's function defined in Eq. (4) and (5), and the hybridization functions read:
\begin{eqnarray}\label{hybridization}
\Delta_{\sigma\sigma'}^1(i \omega_n) & \equiv & \sum_l\Big(\frac{V_{\sigma,l}V_{\sigma',l}}{\epsilon_l-i\omega_n} + \frac{W_{\sigma,l}W_{\sigma',l}}{\epsilon_l+i\omega_n}\Big) \nonumber \\
\Delta_{\sigma\sigma'}^2(i \omega_n) & \equiv &  \sum_l\Big(\frac{V_{\sigma,l}W_{\sigma',l}}{\epsilon_l-i\omega_n} +
\frac{W_{\sigma,l}V_{\sigma',l}}{\epsilon_l+i\omega_n}\Big).
\end{eqnarray}

Hence, we obtain a set of local self-energies $\Sigma^{(i)}_{{\rm imp}, \sigma\sigma'}(i\omega_n)$,
\begin{eqnarray}
\mathbf{\Sigma}_{\rm imp, \sigma\sigma'}(i\omega_n)  =  (i\omega_n\sigma_z + \mu_\sigma)\delta_{\sigma\sigma'} +
\boldsymbol{\Delta}_{\sigma\sigma'} (i\omega_n) - \mathbf{G}^{-1}_{\rm imp, \sigma\sigma'}(i\omega_n).
\label{ssG}
\end{eqnarray}
Then we employ the Dyson equation in real-space representation in order to compute the interacting
lattice Green's function
\begin{equation} \label{G}
 \mathbf{G}(i\omega_n)^{-1}=\mathbf{G}_0(i\omega_n)^{-1}-\mathbf{\Sigma}(i\omega_n).
\end{equation}
The site-dependence of the Green's functions is shown by boldface
quantities that denote a matrix form with site-indexed elements.
Here $\mathbf{G}_0(i\omega_n)^{-1}$ stands for the inverse non-interacting
Green's function
\begin{equation}\label{G0}
 \mathbf{G}_0(i\omega_n)^{-1}=(\mu+i\omega_n)\mathbf{1}-\mathbf{t}.
\end{equation}
In this expression, $\mathbf {1}$ is the unit matrix, the matrix elements $t_{ij}$ are hopping
amplitudes for a given lattice structure.
Eventually the self-consistency loop is closed by specifying the
Weiss Green's function via the local Dyson equation
\begin{equation}\label{g0}
 \Big(\boldsymbol{\mathcal{G}}^{(i)}_{0,\sigma\sigma^\prime}(i\omega_n)\Big)^{-1}=\Big( \mathbf{G}^{(i)}_{\sigma\sigma^\prime}(i\omega_n)\Big)^{-1} + \mathbf{\Sigma}^{(i)}_{\sigma\sigma^\prime}(i\omega_n),
\end{equation}
where the diagonal elements of the lattice Green's function yield
the interacting local Green's function $
\mathbf{G}^{(i)}_{\sigma\sigma^\prime}(i\omega_n)=
(\mathbf{G}_{\sigma,\sigma^\prime}(i\omega_n))_{ii} $. This
self-consistency loop is repeated until the desired accuracy for
superfluid order parameters and Anderson parameters is obtained.

\subsection{Energy within RBDMFT}
\begin{figure*}[t]
\hspace{-25pt}
\centering
\includegraphics*[width=\linewidth]{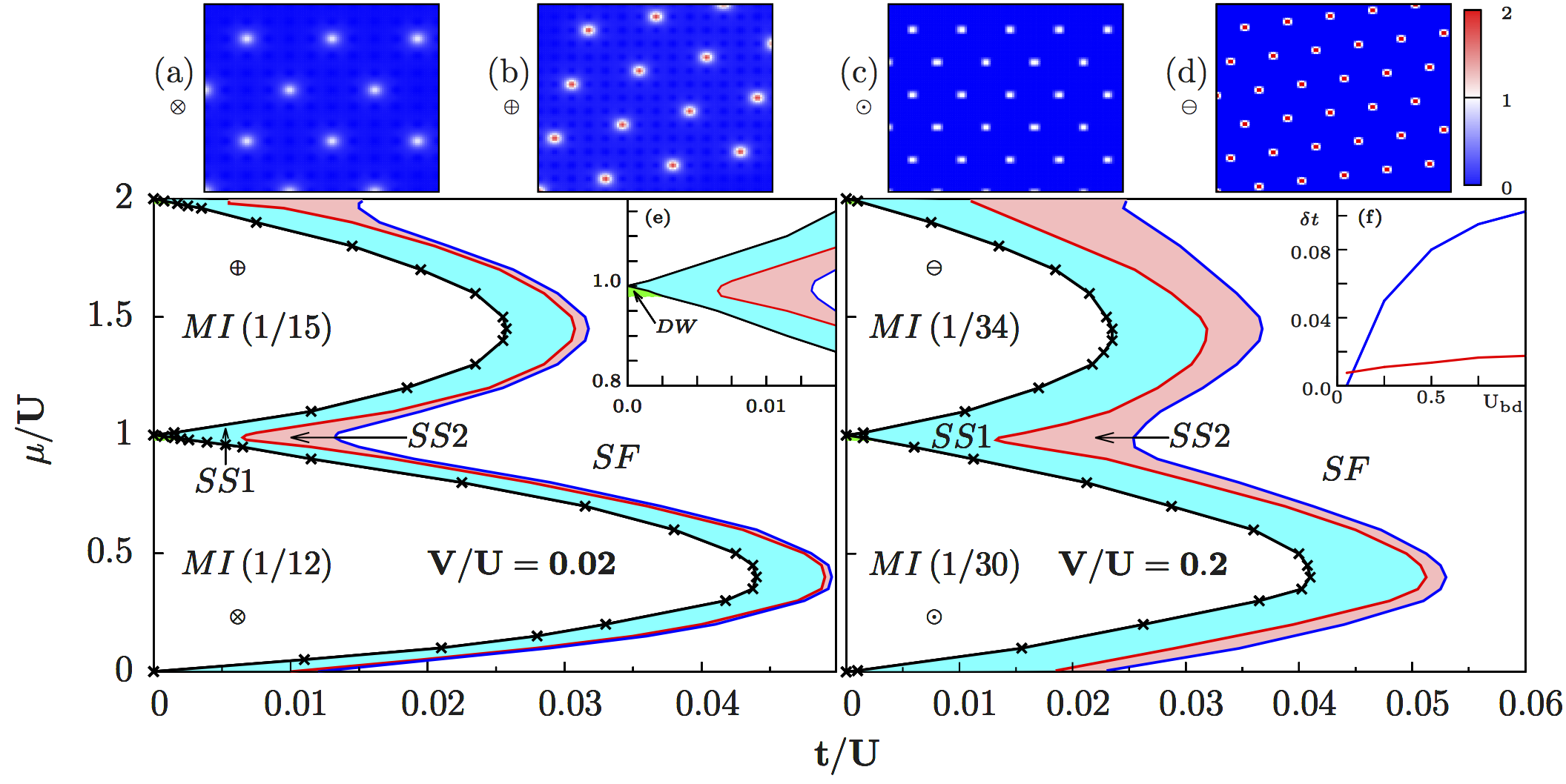}
\vspace{-5pt}
\caption{(Color online) Phase diagram on a square lattice for Rydberg dressed interaction $V/U=0.02$ and $0.2$, respectively, demonstrating stable supersolid regions marked by the cyan (SS1) and pink color (SS2). In the Mott-insulating phase (MI) with spatially uniform total density, the Rydberg dressed species exhibits different crystalline order, as shown in (a)-(d) for real-space density $n_{d}$, with lattice sizes being the square of the area of the unit cell of the Rydberg dressed species [$N_{\rm lat} = 12\times12 \,(\otimes)$; $N_{\rm lat} = 15\times15 \,(\oplus)$; $N_{\rm lat} = 30\times30 \,(\odot)$; and $N_{\rm lat} = 34\times34\,(\ominus)$]. Inset: density-wave phase (DW) with density $n_b=1$ for the ground-state species and $n_d=2$ for the Rydberg dressed state in the corresponding filled sites, respectively (e), and width of supersolid phase [SS1 (blue) and SS2 (red)] $\delta t \equiv t_{\rm c1}-t_{\rm c2}$ as a function of interspecies interaction $U_{bd}/U$ for Rydberg dressed interaction $V/U=0.2$ and chemical potential $\mu /U=0.4$ (f). Other parameters are $U_{bd}=U$, $\mu/U=\mu_{b}/U =\mu_d/U-0.05$. }\label{phase_transition}
\end{figure*}
Calculation of energy is not straightforward within RBDMFT, since the kinetic energy kinetic is given in terms of
non-local expectation values. It can be shown that within the RBDMFT self-consistency conditions, kinetic energy can also be written in terms of Anderson impurity hybridization functions and local Green's functions. A detailed derivation can be found in Ref.~\cite{Huber}.
\subsubsection{Kinetic energy}
In terms of creation and annihilation operators for bosons, $b^\dagger_{i\sigma}$ and $b_{i\sigma}$, respectively, kinetic energy has the form
\begin{equation}\label{Ham}
\hat{H}_{\rm kin} = -\sum_{\langle ij \rangle,\sigma} t_\sigma(b^\dagger_{i,\sigma}b_{j,\sigma}  +  H.c. ).
\end{equation}

Thus expressing the total kinetic energy in terms of real-space Green's functions yields
\begin{align}
E_{\rm kin} &= - \sum_{ij,\sigma} t_{ij}^{\sigma} \langle \hat{b}^{\dagger}_{\sigma,i} \hat{b}_{\sigma,j} \rangle \\
&= \sum_{ij,\sigma} t_{ij}^{\sigma} \left( \lim_{\epsilon \rightarrow 0^{+}} \sum_{n=-\infty}^{\infty} \frac{e^{i\omega_n \epsilon}}{\beta} G_{ji,\sigma}(i\omega_n)  - \phi^*_{i,\sigma}\phi_{j,\sigma} \right)
\end{align}
This expression can be further simplified by employing both the local and lattice Dyson equations within RBDMFT
\begin{align}
\mathbf{G}^{C}_i(i\omega_n)^{-1} &= i\omega_n \sigma_z + \mu + \mathbf{\Delta}_i(i\omega_n) - \mathbf{\Sigma}_i(i\omega_n) \label{eq:local Dyson} \\
[\mathbf{G}^{C}_R(i\omega_n)^{-1}]_{ij} &= t_{ij}\mathbf{1} + \delta_{ij}(i\omega_n \sigma_z + \mu\mathbf{1} - \mathbf{\Sigma}_i(i\omega_n)), \label{eq:lattice Dyson}
\end{align}
which yields
\begin{align}
\sum_j [\mathbf{G}^{C}_R(i\omega_n)^{-1}]_{ij} [\mathbf{G}(i\omega_n)]_{ji} = \sum_j \left[ t_{ij}\mathbf{1}_2 - \delta_{ij}\left( \mathbf{\Delta}_i(i \omega_n)- \mathbf{G}_i(i\omega_n)^{-1} \right) \right] [\mathbf{G}(i\omega_n)]_{ji}
\end{align}
Further using the self-consistency property of the impurity Green's function leads to
\begin{align}
\sum_j t_{ij}[\mathbf{G}(i\omega_n)]_{ji} =  \mathbf{\Delta}_i(i\omega_n) \mathbf{G}_i(i\omega_n).
\end{align}
and we finally obtain
\begin{align}
E_{\rm kin} = & \frac{2}{\beta} \lim_{\epsilon \rightarrow 0^{+}} \sum_{i\sigma n\geq0} \textrm{Re}\left( \left[ \mathbf{\Delta}_{\sigma,i}(i\omega_n) \mathbf{G}_{\sigma,i}(i\omega_n)\right]_{11} e^{i\omega_n \epsilon} \right) \nonumber \\ & - \sum_{ij\sigma} t_{ij}\phi^*_{i,\sigma}\phi_{j,\sigma} - \frac{\textrm{Tr} \left[ \mathbf{\Delta}_{\sigma,i}(0) \mathbf{G}_{\sigma,i}(0)\right]}{2 \beta}.
\end{align}

\subsubsection{Total energy}
The ground state within RBDMFT corresponds to the solution with the lowest energy, where the corresponding total energy of the impurity site which is given as follows:
\begin{eqnarray}
E=E_{\rm kin}+E_{\rm int}.
\end{eqnarray}
For the Bose-Hubbard model of spin-1 bosons, the on-site interaction term is given by:
\begin{eqnarray}
E_{\rm int}=\frac12 \sum_{i,\sigma\sigma^\prime} U_{\sigma\sigma^\prime} \hat{n}_{i,\sigma}(\hat{n}_{i,\sigma^\prime}-\delta_{\sigma\sigma^\prime}) + \sum_{i<j}V_{ij}\hat{n}_{i,d}\hat{n}_{j,d}.
\end{eqnarray}

\section{Numerical results within RBDMFT}
\subsection{Density dependent phase diagram of Rydberg-dressed systems}
\begin{figure*}[t]
\hspace{-25pt}
\centering
\includegraphics*[width=\linewidth]{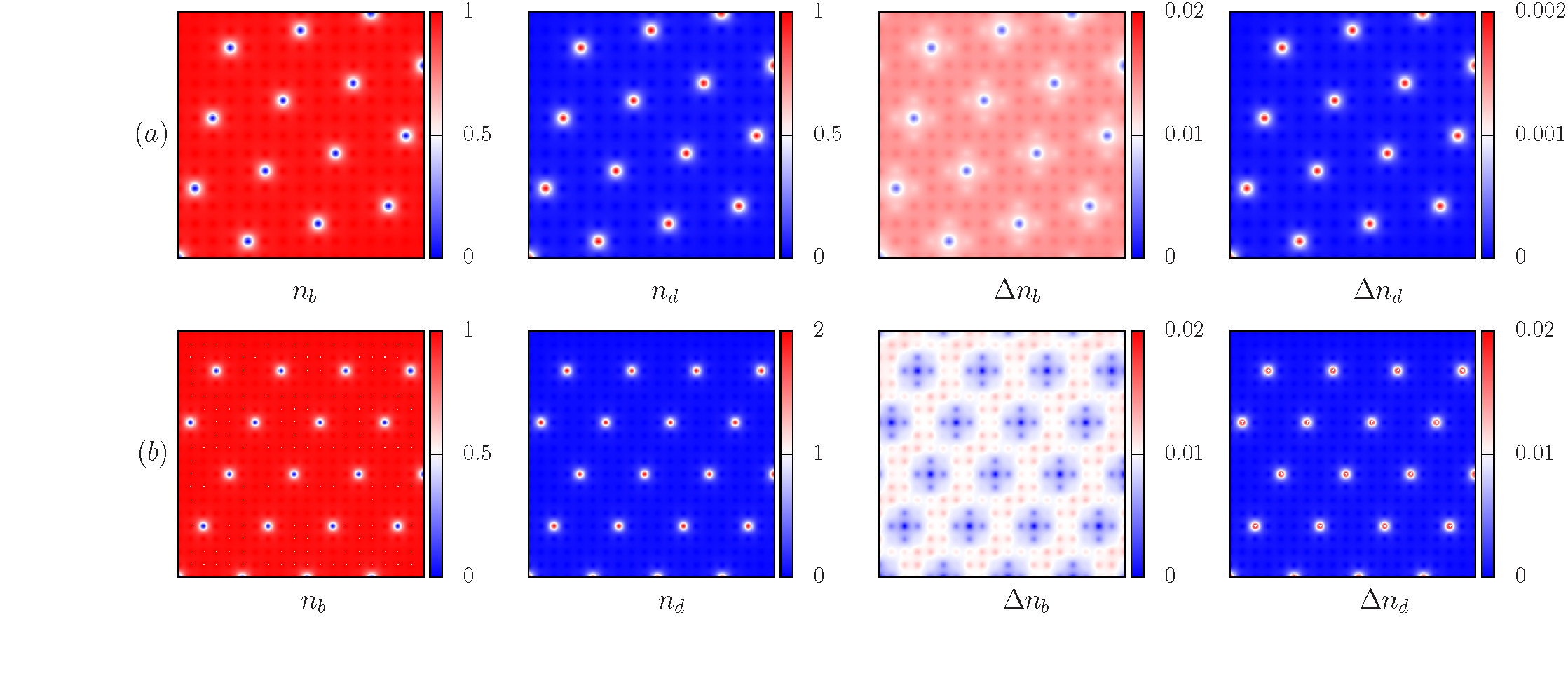}
\vspace{-20pt}
\caption{(Color online) Real-space density $n_{b,d}$ and density fluctuations $\Delta_{b,d}\equiv (n_{b,d} - \langle n_{b,d} \rangle)^2$ in MI (a) and CDW (b) phases, with lattice sizes being the square of the area of the unit cell of the Rydberg dressed species [$N_{\rm lat} = 15\times15$ (a)] and [$N_{\rm lat} = 20\times20$ (b)], respectively. Other parameters are $t/U=0.03$, $V/U=0.3$, $U_{bd}=U$, $\mu_b/U=0.2$ and $\mu_d/U=0.7$ (a), and $t/U=0.0023$, $V/U=0.02$, $U_{bd}=U$, $\mu_b/U=0.98$ and $\mu_d/U=1.03$ (b) (see Fig.1 in the supplementary).}\label{fluctuation}
\end{figure*}
In this paragraph, we study the stability of quantum phases of Rydbery-dressed systems in optical lattices for different fillings. In the strong coupling limit with $U_{\sigma\sigma^\prime} \gg t$, we find that the system favors Mott insulating or density-wave phase with different types of crystalline order in the individual species. Interestingly, we observe a density-wave phase with a nonuniform total density which breaks lattice translational symmetry, with densities $n_{ib}=1$ and $n_{id}=2$, appears, as shown in green region of Fig.~\ref{phase_transition}. These density waves exhibit nonzero density fluctuations, as shown in Fig.~\ref{fluctuation}. However, quantum fluctuations as a result of higher-order tunneling processes are weak, due to the strong long-range interactions. Actually, the density wave of the dressed species is also predicted in the single-species case~\cite{long-range}.

Away from the deep MI regime, i.e. in the intermediate hopping regime, we observe two types of quantum phase transition from MI to supersolid, i.e. the uncoupled ground-state species demonstrates a phase transition from MI to supersolid, and then followed by the Rydberg dressed species, as shown in the Fig.~\ref{phase_transition}. Interestingly, we observe a pronounced region of supersolid appearing in our simulations, as a result of the onsite interspecies interactions, indicating a higher chance for directly observing these phases in realistic experiments, compared to single-species case~\cite{long-range}. Actually, we indeed observe the width of SS1 and SS2 shrinks as a function of interspecies interactions, as shown in Fig.~\ref{phase_transition}(f), where SS1 clearly disappears for smaller $U_{bd}$. In addition, the long-range interaction also shifts the phase transition between MI and SS1, even though the bare species only possess onsite interactions. As shown in Fig. 2 in the main text, the phase boundary shrinks to lower hopping regime with increasing the long-range interaction $V$.

Finally, in the weakly interacting regime with $t \gg U_{\sigma\sigma^\prime}$, a superfluid phase with uniform total density distribution is found in our simulations, where both species demonstrate homogenous density distribution. Here, crystalline orders are destroyed by the large density fluctuations, and the system only supports superfluidity with uniform density.

\subsection{Dipolar system}
\begin{figure}[h]
\includegraphics*[width=5in]{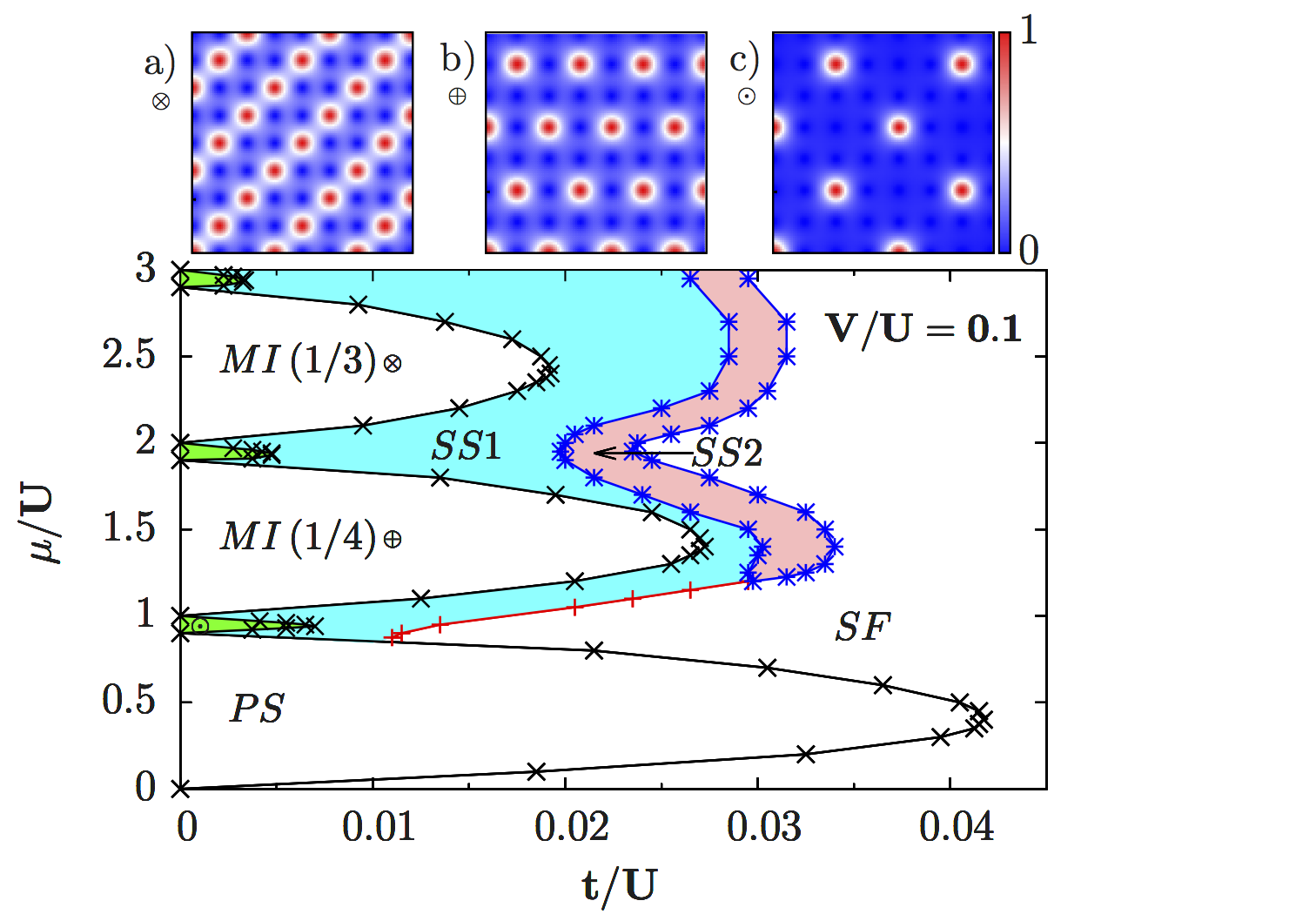}
\vspace{-5mm}
\caption{(Color online) Phase diagram for a mixture of nondipolar species $b$ and dipolar component $d$ on a square
lattice for a dipolar interaction strength $V/U=0.1$, exhibiting pronounced regions of supersolid marked by the cyan (SS1) and pink color (SS2). In contrast to nearest-neighbor case~\cite{NN}, the system demonstrates various crystalline order, as shown in a)-c) for the real-space density distribution of the dipolar species. Note here that, in the DW, marked by the green color, the total density distributes spatially nonuniform with a homogeneous density for the nondipolar species, whereas, in the MI, the total density distribute spatially uniform. We observe a phase separation (PS) in the MI region with a total filling $n_b+n_d=1$, in addition to spatially uniform superfluid (SF). Other parameters are $U_{bd}=0.9U$, and $\mu_{b,d}=\mu$.}\label{density_trap}
\end{figure}
We have so far studied crystalline order in the Rydberg dressed systems. Actually, the physics of these competing orders can also be exhibited in dipolar system loaded in an optical lattice, along with quick developments in the cooling and trapping of magnetic atoms~\cite{Cy_Dr_Ey} and diatomic molecules~\cite{Ye}. Recently, a Gutzwiller mean-field phase diagram of a binary Bose mixture on a square optical lattice is studied, where one species possesses a non-negligible dipole moment~\cite{NN}. In their study, only the nearest-neighbor part of the dipolar interactions was included. To obtain a better understanding of the Rydberg dressed system studied above and make a comparison, we here study a mixture of dipolar and nondipolar bosons on a square optical lattice, with {\it real} long-range interactions beyond nearest-neighbor approximations. We study the system by means of RBDMFT, which takes into account quantum fluctuations and is actually a higher-order expansions of Gutzwiller mean-field theory.

In Fig.~\ref{density_trap}, we show the resulting phase diagram of dipolar and nondipolar bosonic mixtures on a 2D optical lattice. In general, there are also five phases in this dipolar system, i.e. SF, MI, DW and two types of supersolid. Compared to nearest-neighbor interaction and static mean-field approximations~\cite{NN}, two big differences have been observed. First, rich crystalline patterns appear in the system, as shown in Fig.~\ref{density_trap}a)-c), with a filling factor of 1/3, 1/4 and 1/8 for the dipolar species, respectively. Second, we observe that the region of supersolid phase is also altered. Note here that we recover the static mean-field phase diagram with nearest-neighbor interactions within Gutzwiller approximations in Ref.~\cite{NN}.

\section{Bogoliubov spectra of the bare species in the SS1 phase}

As the dressed atoms are in a density wave state, we could decouple the total wave function in the ground state as $|\Psi_{\rm{g}}\rangle\approx |{\rm{DW}}_d\rangle \otimes |\Psi_b\rangle$, where $|{\rm{DW}}_d\rangle$ and $|\Psi_b\rangle$ are the wave function of dressed atoms and bare component, respectively. Here quantum fluctuations of the density wave could be neglected. After tracing out the dressed atoms, we obtain an effective Hamiltonian for the bare species,
\begin{eqnarray}
\hat{H}_{\rm{e}}= &-& \sum_{\langle ij \rangle}t(\hat{b}^\dagger_{i}\hat{b}_{j} + {\rm H.c.})
+\frac{U}{2} \sum_{i} \hat{n}_{i}(\hat{n}_{i}-1)
-  \sum_{i} \mu \hat{n}_{i}+U_1\sum_{\{j\}} \hat{n}_{j},
\label{eq:eff}
\end{eqnarray}
where ${\{j\}}$ denotes sites of the oblique lattice occupied by dressed atoms with the corresponding particle number $n_d$. For parameters considered in this work, numerical results show that $n_d\approx 1$. Through Fourier transformation, we can derive the Hamiltonian in momentum space in the first Brillouin zone,
\begin{eqnarray}
\tilde{H}=-\sum_{{\vec{k}}}[\mu+2t(\cos k_xa + \cos k_ya)]b_{\vec{k}}^{\dagger}b_{\vec{k}}
+ \frac{U}{2N}\sum_{{\vec{k}}_1{\vec{k}}_2{\vec{k}}_3}b^{\dagger}_{{\vec{k}}_1}b^{\dagger}_{{\vec{k}}_3}b_{{\vec{k}}_3+{\vec{k}}_2}b_{{\vec{k}}_1-{\vec{k}}_2} +U_1\sum_{\{{\vec{k}}\}}b^{\dagger}_{\vec{k}}b_{\vec{k}},
\label{eq:effK}
\end{eqnarray}
where $N$ is the total number of sites and $U_1=\bar{n}_d U_{bd}$ with $\bar{n}_d = n_dN_d/N$. $N_d$ is the number of sites occupied by the dressed atoms, and $\{k\}$ denotes momentum spanned in the first Brillouin zone of the lattice occupied by the dressed atoms.

Expanding the Hamiltonian~(\ref{eq:effK}) around $|\vec{k}|=0$ and keeping only quadratic terms of the operators, this yields,
\begin{eqnarray}
\tilde{H}\approx E_0 - \sum_{{\vec{k}}\neq 0}\left[\mu+2t(\cos k_xa + \cos k_ya)-2U\bar{n}_b\right]b_{\vec{k}}^{\dagger}b_{\vec{k}}
+ \frac{U\bar{n}_b}{2}\sum_{{\vec{k}}\neq 0}(b_{\vec{k}}b_{-{\vec{k}}}+b^{\dagger}_{-{\vec{k}}}b^{\dagger}_{\vec{k}})+U_1\sum_{\{{\vec{k}}\neq 0\}}b^{\dagger}_{\vec{k}}b_{\vec{k}},
\label{eq:expand}
\end{eqnarray}
where $E_0=-UN_0^2/2N$ is the energy of the condensed atoms, with $N_0$ to be the number of condensed atoms and $\mu=-4t + U\bar{n}_b  +U_1$ the chemical potential and the mean occupation of the condensed atom $\bar{n}_b=N_0/N$.

As the interspecies interaction [the last term in Eq.~(\ref{eq:expand})] only appears in the low momentum regions (Brillouin zone $\{k\}$), we will have two different forms of the approximate Hamiltonian depending on values of the momentum. Substituting the chemical potential $\mu$, we get the approximate Hamiltonian within the first Brillouin zone of the dressed atom,
\begin{eqnarray}
\tilde{H}\approx E_0 + \sum_{{\vec{k}}\neq 0}\left[\varepsilon_k+U \bar{n}_b\right]b_{\vec{k}}^{\dagger}b_{\vec{k}}
+\frac{U\bar{n}_b}{2}\sum_{{\vec{k}}\neq 0}(b_{\vec{k}}b_{-{\vec{k}}}+b^{\dagger}_{-{\vec{k}}}b^{\dagger}_{\vec{k}}),
\label{eq:inbrillouin}
\end{eqnarray}
and the corresponding Bogoliubov spectrum is
\begin{eqnarray}
E_l(k)=\sqrt{\varepsilon_k(\varepsilon_k + 2U\bar{n}_b)},
\label{eq:ispectra}
\end{eqnarray}
with $\varepsilon_k = -2t(\cos k_x a + \cos k_y a -2)$. The spectrum is similar to the one of a weakly interacting Bose gas in a square optical lattices.

For momenta outside the first Brillouin zone of the dressed atoms, we have a different form of the approximate Hamiltonian,
\begin{eqnarray}
\tilde{H}\approx E_0 + \sum_{{\vec{k}}\neq 0}\left[\varepsilon_k+U\bar{n}_b-U_1\right]b_{\vec{k}}^{\dagger}b_{\vec{k}}
+\frac{U\bar{n}_b}{2}\sum_{{\vec{k}}\neq 0}(b_{\vec{k}}b_{-{\vec{k}}}+b^{\dagger}_{-{\vec{k}}}b^{\dagger}_{\vec{k}}),
\label{eq:outbrillouin}
\end{eqnarray}
the corresponding Bogoliubov spectra is
\begin{eqnarray}
E_l(k)=\sqrt{(\varepsilon_k-U_1)(\varepsilon_k -U_1 + 2U\bar{n}_b)},
\label{eq:ospectra}
\end{eqnarray}
which will be nonzero only at large momentum (outside the first Brillouin zone).

The roton instability occurs at the boundary of the two Bogoliubov spectrum. Using Eq.~(\ref{eq:ospectra}), we can find the spectrum becomes complex when $\varepsilon_k<U_1$. This allows us to find the critical value of the tunneling rate $t_c$
\begin{eqnarray}
t_c = \frac{U_1}{2[2-\cos k^{(b)}_x -\cos k^{(b)}_y]},
\end{eqnarray}
where $k^{(b)}_x$ and $k_y^{(b)}$ are values of the momentum at the boundary of the first Brillouin zone of the oblique lattice.

The soft-core interaction will affect structures of the oblique lattice. Therefore the critical $t_c$ will change as the interaction $V$ changes. As shown in Fig. 3 in the main text, the first Brillouin zone is not of a regular shape, such that the critical value $t_c$ will vary with both $k^{(b)}_x$ and $k^{(b)}_y$. To show this we evaluate the critical values $t_c$ using the crystalline structure of the dressed atoms at the SS1-SS2 phase boundary, which are  obtained by the full numerical calculation.  For example, $t_c$ lies in a range $[0.087,0.094]$ if $V=0.3$. When the long range interaction becomes strong, we find that the range of critical $t_c$ increases. For example, $t_c\in [0.085,0.11]$ when $V=0.4$,  and $t_c\in [0.073,0.13]$ when $V=0.6$. Although these values are close to the numerical calculations, it is apparent that one will not be able to determine phase boundaries accurately using the Bogoliubov calculation.

Another limitation of this calculation is that areas of the crystalline structure become smaller when $V$ is weak. Long range correlations become important in determining the ground state phases, which prevents us to decouple the total wave function into two parts. In this regime, the Bogoliubov calculation fails to capture the many body physics.

\end{widetext}
\end{document}